\documentclass[review]{elsarticle}

\usepackage{lineno,hyperref}
\usepackage{multirow}
\usepackage{amsmath}
\usepackage{caption}
\usepackage[utf8]{inputenc}
\usepackage{verbatim}
\usepackage{longtable}
\usepackage[colorinlistoftodos]{todonotes}
\usepackage{siunitx}
\usepackage{isotope}
\usepackage{xcolor}
\usepackage{booktabs} 

\newcommand{\eq}{\; = \;}

\modulolinenumbers[5]

\journal{Applied Radiation and Isotopes}

\begin{document}

\begin{frontmatter}

\title{Methodology for measuring photonuclear reaction cross sections with an electron accelerator \\ based on Bayesian analysis}

\author[lhep]{Saverio~Braccini}
\author[lhep]{Pierluigi~Casolaro\corref{cor}}
\author[lhep]{Gaia~Dellepiane}
\author[metas]{Christian~Kottler}
\author[metas]{Matthias~L\"uthi\corref{cor}}
\author[lhep,insel]{Lorenzo~Mercolli}
\author[metas]{Peter~Peier}
\author[lhep,napoli]{Paola~Scampoli}
\author[dcb]{Andreas~T\"urler}

\affiliation[lhep]{ organization={Albert Einstein Center for Fundamental Physics (AEC), Laboratory for High Energy Physics (LHEP), University of Bern},
                    postcode={3012},
                    city={Bern},
                    country={Switzerland}}               
\affiliation[metas]{organization={Federal Institute of Metrology METAS},
                    postcode={3003},
                    city={Bern-Wabern},
                    country={Switzerland}}
\affiliation[insel]{organization={Department of Nuclear Medicine, Inselspital, Bern University Hospital, University of Bern},
                    postcode={3010},
                    city={Bern},
                    country={Switzerland}}
\affiliation[dcb]{organization={Department of Chemistry, Biochemistry, and Pharmaceutical Sciences (DCBP), University of Bern},
                    postcode={3012},
                    city={Bern},
                    country={Switzerland}}
\affiliation[napoli]{organization={Department of Physics “Ettore Pancini", University of Napoli Federico II, Complesso Universitario di Monte S. Angelo},
                    postcode={80126},
                    city={Napoli},
                    country={Italy}}

\cortext[cor]{Corresponding authors: pierluigi.casolaro@unibe.ch, matthias.luethi@metas.ch}

\begin{abstract}
Accurate measurements of photonuclear reaction cross sections are crucial for a number of applications, including radiation shielding design, absorbed dose calculations, reactor physics and engineering, nuclear safeguard and inspection, astrophysics, and nuclear medicine. Primarily motivated by the study of the production of selected radionuclides with high-energy photon beams (mainly \isotope[225]{Ac}, \isotope[47]{Sc}, and \isotope[67]{Cu}), we have established a methodology for the measurement of photonuclear reaction cross sections with the microtron accelerator available at the Swiss Federal Institute of Metrology (METAS). The proposed methodology is based on the measurement of the produced activity with a High Purity Germanium (HPGe) spectrometer and on the knowledge of the photon fluence spectrum through Monte Carlo simulations. The data analysis is performed by applying a Bayesian fitting procedure to the experimental data and by assuming a functional trend of the cross section, in our case a Breit-Wigner function. We validated the entire methodology by measuring a well-established photonuclear cross section, namely the \isotope[197]{Au}($\gamma$, n)\isotope[196]{Au} reaction. The results are consistent with those reported in the literature.
\end{abstract}


\end{frontmatter}

\section{Introduction}
Photonuclear reactions occur when megaelectronvolt photons undergo an inelastic interaction with a nucleus. At photon energies below 25-30 MeV, the excitation function of photonuclear reactions is characterized by a prominent peak, known as giant dipole resonance (GDR), that is a collective excitation of the atomic nucleus in which nucleons move together to create a large oscillation of the nucleus in the shape of a dipole.
This energy range matches the upper limit of most electron accelerators, which produce X-rays by "Bremsstrahlung", i.e. by slowing down (or stopping completely, depending on the thickness of the target) the electrons in a target. The photon flux scales approximately quadratically with the target atomic number, thus high Z materials are typically chosen as converter targets. The most commonly used are gold, tantalum or tungsten, although the use of lighter materials such as niobium and copper is also reported \cite{zilges2022photonuclear}. Experimentally, measuring photonuclear cross sections at bremsstrahlung facilities is challenging. The reaction yield is a folding of the cross section and of the continuous X-ray energy spectrum, and the yield curve can be obtained experimentally by varying the electron energy. Thus, the cross section curve is typically evaluated by means of unfolding methods. Of course, this relies heavily on the knowledge of the energy spectrum (which is difficult to measure), reproducibility of the accelerator output, and high counting statistics. In order to circumvent the issues of a bremsstrahlung spectrum, the production of high-energy X-rays has been also achieved with other techniques including Laser‐Compton Scattering (LCS). While bremsstrahlung generally produces a larger number of photons, LCS has the advantage of producing quasi-monochromatic gamma rays, which allow to avoid the use of unfolding methods \cite{turturica2019investigation}. In spite of these difficulties, now there is a large amount of measured data available from photonuclear reactions. Along this line, the International Atomic Energy Agency (IAEA) issued a comprehensive review on photonuclear data, emphasizing the importance of the accurate knowledge of photonuclear reaction cross sections for several applications \cite{kawano2020iaea}, including radiation shielding design and transport analyses, calculation of the absorbed dose in human body for radiotherapy, physics and technology of fission and fusion reactors, activation analyses, safeguards and inspection technologies, nuclear waste transmutation and astrophysical nucleosynthesis. 

In the last decade, the possibility of using photonuclear reactions for the production of radionuclides for nuclear medicine has been established \cite{Maslov2006, rane201547ca, mamtimin2015sc, Starovoitova2014}. The renewed interest in this topic was sparked by the commercial availability of compact high-power electron accelerators, such as the 35 MeV, 120 kW linac from MEVEX Corp (Stittville, ON, Canada) and the Rhodotron TT300-HE, an electron accelerator characterized by a maximum energy of 40 MeV and a beam power up to 125 kW, produced by IBA (Louvain-La-Neuve, Belgium).
Of course, the precise knowledge of interaction cross sections is key for a scalable production of radionuclides for medical purposes. At the Bern University Hospital's medical cyclotron facility, cross sections of several proton-induced nuclear reactions were measured, in particular those involving the production of so-called theranostic pairs, such as \isotope[43,44]{Sc}/\isotope[47]{Sc}, \isotope[61,64]{Cu}/\isotope[67]{Cu} and \isotope[152,155]{Tb}/\isotope[149,161]{Tb}, as well as more recently of the Auger emitter \isotope[165]{Er} that can be potentially be used in combination with other lanthanides \cite{dellepiane202247sc, dellepiane2022cross, dellepiane2022cross1, dellepiane2022cross12, dellepiane2022cross2, dellepiane2023cross}. Currently, we are investigating the feasibility of the METAS electron microtron (maximum energy: 22 MeV, average current: 20 $\mu$A) for studying selected photonuclear reactions, in particular for the production of
 \isotope[225]{Ac} $ \left[\mathrm{t}_{1 / 2}=9.9 \mathrm{~d}, \mathrm{E}_{\alpha}=5.8 \mathrm{~MeV}(100 \%) \right]$,  \isotope[47]{Sc} $ [\text{t}_{1/2}=3.349 \text{ d}, \text{E}_{\beta^{-}}^{\text{max}}=440.9 \text{ keV} (68.4\%);$
$600.3 \text{ keV} (31.6\%)$, $ \text{E}_\gamma=159.4 \text{ keV} (68.3\%)]$,
and  \isotope[67]{Cu} $ [ E_{\beta-}^{\max }=377 \mathrm{~keV} (57 \%) ; 468 \mathrm{~keV} (22\%) ; 562 \mathrm{~keV} (20 \%)]$.

In particular, \isotope[225]{Ac} is one of the most promising radionuclides for Targeted Alpha Therapy (TAT). Recent findings have demonstrated the striking potential of \isotope[225]{Ac}-PSMA-617 for prostate cancer therapy \cite{kratochwil2020225ac}. To date, the availability of \isotope[225]{Ac} is still insufficient with respect to the high demand for clinical applications. The main production routes are the radiochemical extraction from \isotope[229]{Th}, high-energy proton induced spallation of \isotope[232]{Th} and \isotope[238]{U} targets \cite{robertson2019design}, and neutron irradiation of \isotope[232]{Th} and \isotope[226]{Ra} targets \cite{hoehr2017medical}. A viable, but not yet fully studied alternative route for the production of \isotope[225]{Ac} in large scale is the irradiation of \isotope[226]{Ra} targets with high-energy gamma rays \cite{melville2007production}. In view of the assessment of the \isotope[226]{Ra}($\gamma$, n)\isotope[225]{Ra} cross section, we aim to establish a rigorous procedure for the measurement of photonuclear reactions at METAS. This paper reports on the validation of this procedure through the measurement of a well-established photonuclear monitor reaction, namely the \isotope[197]Au$(\gamma, n)$\isotope[196]{Au} reaction \cite{thiep2006experimental, veyssiere1970photoneutron, itoh2011}. The Methods section describes the microtron accelerator at METAS, irradiation and measurements, Monte Carlo simulations, and data analysis. The results are presented and discussed in the following two sections, and eventually conclusion and outlook are drawn.

\section{Materials and methods}\label{s:methods}
After describing the electron accelerator at METAS in the first paragraph, the irradiation procedures and the measurements with gamma spectroscopy are discussed in the second paragraph. The third paragraph focuses on the assessment of the photon fluence spectrum through Monte Carlo simulation. Finally, the data analysis techniques are discussed.

\subsection{The accelerator at METAS}\label{s:microtron}

The irradiation experiments were conducted at the electron accelerator of the Swiss Federal Institute of Metrology (METAS). The accelerator is of microtron type, capable of producing electron beams with an endpoint energy from \SIrange{4}{22}{\mega\electronvolt}. The installed accelerator is based on the design described in Ref.~\cite{Svensson1977}.

\begin{figure}
  \centering 
	\hfill 
  \includegraphics[width=1\textwidth]{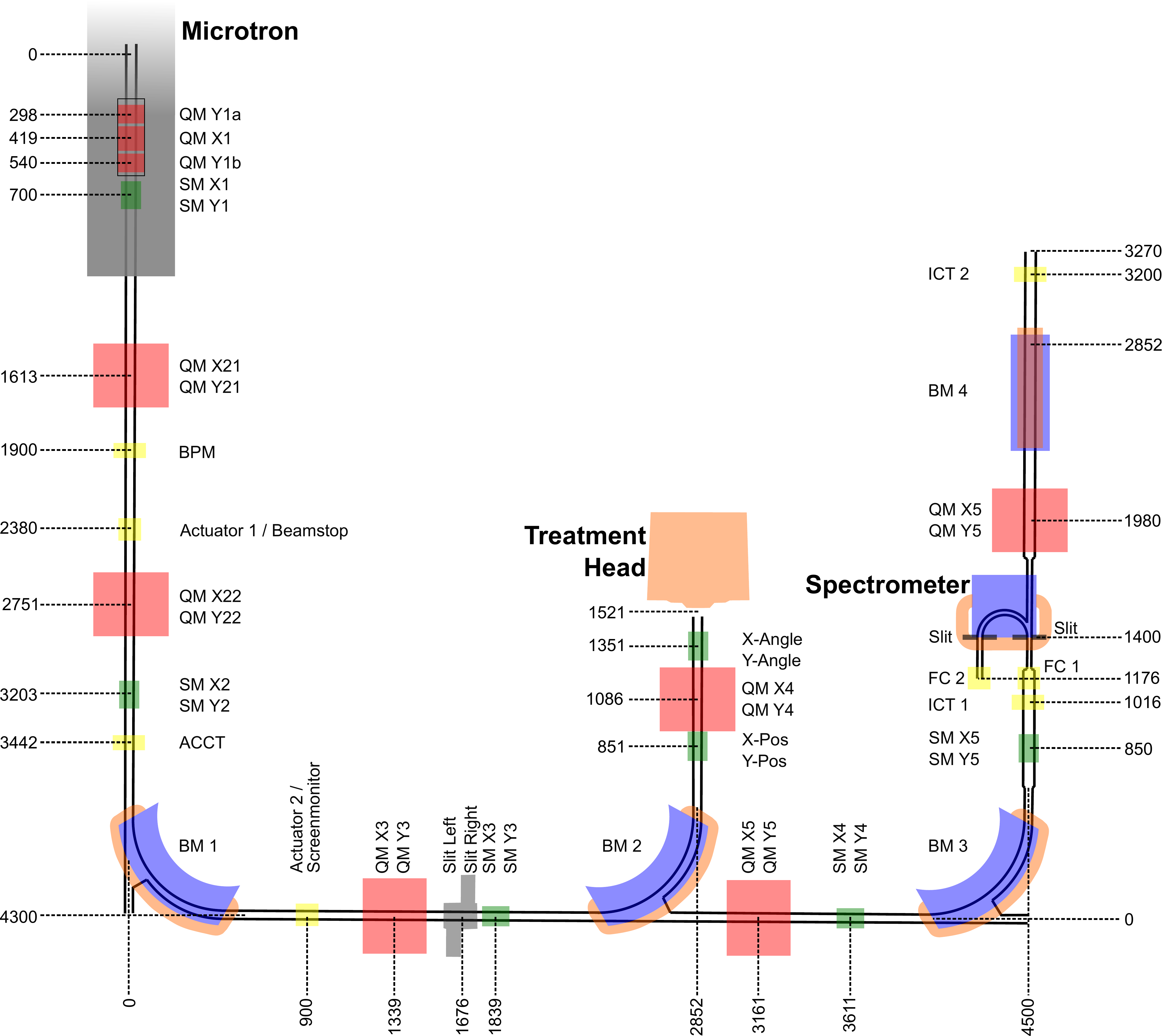}
  \caption{Scheme of the main elements of the Microtron accelerator facility at METAS. Quadrupole magnets are depicted in red, steering magnets in green and bending magnets in blue. Instrumentation in shown in yellow.} \label{f:scheme_Microtron}
\end{figure}

The relevant parts of the accelerator facility are shown in Fig~\ref{f:scheme_Microtron}. The initial electron beam is formed in an electron gun and is accelerated inside a resonator (\SI{535}{\kilo\electronvolt} per revolution). The electron beam cycles through the resonator, by means of a constant magnetic field, until its path is offset by the extraction tube. Subsequently, the electron beam enters the beamline. Here the beam is shaped by means of four quadrupole magnets (QM1 to 4) and four steering magnets (SM1 to 4), and transported via two bending magnets (BM1 and 2) to the treatment head. The beam is extracted from the vacuum tube to air through a \SI{400}{\micro\meter} thick aluminum window and directed onto a converter target. A gold plate (\SI{2}{\milli\meter} thick with a diameter of \SI{10}{\milli\meter}) acts as a Bremsstrahlung converter. Thermal cooling of the converter is provided by a copper housing, through which water is circulated. Water and copper located behind the gold disc (in the beam direction) absorb residual emerging electrons and low energy photons, hence hardening the photon beam. A tungsten block with a conical opening acts as a collimator. Under normal operation conditions, the photon energy spectrum would be further shaped by a flattening filter located downstream of the collimator. The filters are interchangeable by a revolver assembly. For our irradiations the flattening filter was replaced by a custom target mount, described in Sec.~\ref{s:irradiations}.

A single electron beam pulse has a duration of \SI{3}{\micro\second} and a current of \SIrange{25}{100}{\milli\ampere} (depending on beam energy). The repetition rate can be varied step-wise in the range of \SIrange[range-phrase={~and~}]{1}{200}{\hertz}. On the converter the beam is assumed to have a Gaussian shape with a full width at half maximum (FWHM) of \SI{3}{\milli\meter}. Although every electron orbit in the accelerator can be accessed by the extraction tube, optimized magnet settings only exist for a subset of orbits. The exact energy corresponding to an orbit, as well as the energy spread within an orbit, was determined using a magnetic spectrometer in a separate beamline, dedicated for total absorption dosimetry \cite{Vrs2012}. The energy spread was found to be of the order of \SI{25}{\kilo\electronvolt} for all measured orbits.

\subsection{Irradiation and measurements}\label{s:irradiations}
\begin{figure}
  \centering 
  \includegraphics[width=1\textwidth]{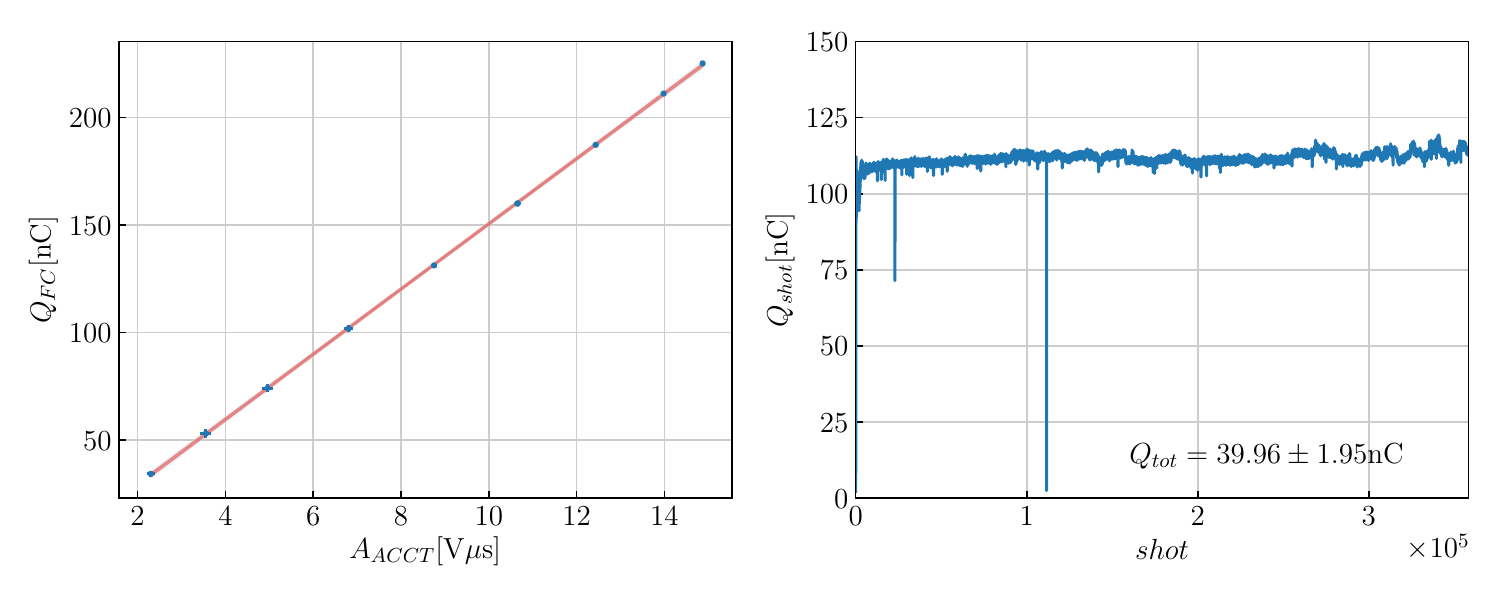}
  \caption{The left hand side show the calibration curve of the recorded ACCT area against the charge collected in the Faraday cup. The right hand side shows the beam charge evolution over the entire irradiation with a beam energy of \SI{20.678}{\mega\electronvolt}} \label{f:beam_charge}
\end{figure}

In order to validate the proposed experimental procedure with the measurement of a well-established photonuclear cross section, we selected the \isotope[197]{Au}($\gamma$, n)\isotope[196]{Au} reaction. Gold foils with a diameter of 25 mm and a nominal thickness of 12.5 $\mu m$ have been purchased from Goodfellow, GmbH. We performed irradiation runs of 11 gold targets at different electron energies in the range 8.499-20.678 MeV. It should be noted that the energy threshold of the \isotope[197]{Au}($\gamma$, n)\isotope[196]{Au} nuclear reaction is $8.070 \pm 0.003 \, \mathrm{MeV}$ \cite{Audi2003}. To evaluate the initial number of target nuclei, each gold foil was weighted prior the irradiation using a precision scale\footnote{Mettler Toledo XP205} with a typical uncertainty of \SI{0.007}{\milli\gram}. Irradiation times were chosen based on the predicted activity and ranged from half an hour (for the irradiation at the highest beam energy) to 6 hours (for the irradiation at the lowest beam energy). The charge of each individual beam pulse was measured using an AC current transformer (ACCT)\footnote{Bergoz ACCT-S-055} connected to a high bandwidth waveform digitizer\footnote{Spectrum Instrumentation, M2p5962-x4}. For each irradiation individual pulses were recorded and summed post-irradiation to obtain the total charge on the Bremsstrahlung converter. This current/charge measurement setup was calibrated against a Faraday Cup. Comparing the simultaneously collected charge in the Faraday cup over a precision resistor (\SI{50}{\ohm}) to the area of the ACCT signal allowed to obtain a linear calibration curve over the range of \SIrange{25}{225}{\nano\coulomb}. Thus, establishing an accurate charge measurement for individual pulses. The calibration curve is shown on the left in Fig \ref{f:beam_charge}, on the right hand side of the figure a typical evolution of the beam current during an irradiation is shown. To investigate the stability of the calibration, the calibration factors were monitored for various beam repetition rates and vertical and horizontal displacements (using steering magnets SM X2 and Y2 until the beam was lost). Based on these investigations the uncertainty of the beam charge was quantified to be within \SI{1.2}{\percent}.

\begin{figure}
  \centering 
	\hfill 
  \includegraphics[width=1\textwidth]{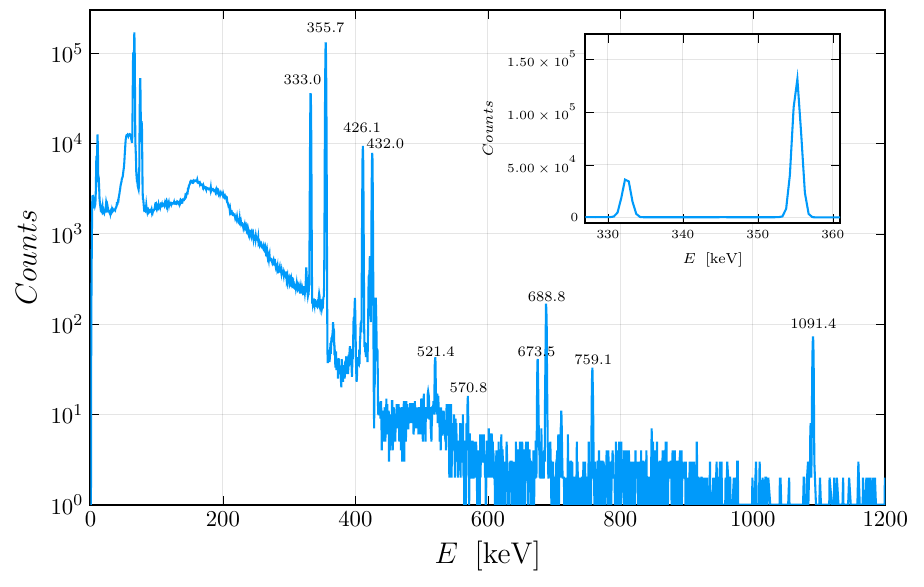}
  \caption{Typical gamma energy spectrum of an irradiated gold target after irradiation.} \label{f:HPGe}
\end{figure}

EBT3 Gafchromic films were used to evaluate the photon beam uniformity in the position of the gold foil \cite{casolaro2019physics}. The beam was found to be uniform within 1\% on the foil surface. The activity of the gold targets at the end of the irradiation was measured with a High Purity Germanium (HPGe) detector in operation at the cyclotron laboratory of the University of Bern. The detector's energy calibration and efficiency are periodically verified with a multi-peak radioactive source type EG 3X from EUROSTANDARD CZ, spol. s r.o., with an energy resolution of 0.24\% (\isotope[137]{Cs} peak FWHM). The detector is used on a daily basis for cross section, activity and half-life measurements of radionuclides of medical interest\cite{dellepiane2021research,juget2023activity,duran2022half}. The detector is a coaxial N-type HPGe (Canberra GR2009) with the sensitive volume shielded by 10 cm of lead. The pre-amplifier signal is fed into a Lynx digital analyzer. The gamma spectra were analyzed using the Interspec software\cite{osti_1833849}. As an example, Fig.~\ref{f:HPGe} shows the gamma spectrum of a gold target after exposure to the Bremsstrahlung beam.

The energies of all the peaks related to the \isotope[196]{Au} decay are highlighted, whereas the inset zooms on the peaks used in the analysis, namely the 355.73 keV (87.0 \%) and the 333.03 keV (22.9 \%). The mass of the gold targets, the irradiation time and the \isotope[196]{Au} activity are reported in Tab. \ref{t:measurements} for all the beam energies.

\begin{table}[]
    \centering
    \begin{tabular}{cccccc}
        \toprule
        $E_{beam} \, [\mathrm{MeV}]$ & $t_{irr} \, [\mathrm{s}]$ & $Q \, [\mathrm{mC}]$ & $t_{dec} \, [\mathrm{s}]$ & $m_{Au} \, [\mathrm{mg}]$ & $A \, [\mathrm{kBq}]$ \\
        \midrule 
        $ 8.499 $ & $ 21660 $ & $ 1130.8 \pm 13.6 $ & $ 25200 $ & $ 123.09 $ & $ (76.0 \pm 8.4 ) \cdot 10^{-3} $ \\
        $ 9.030 $ & $ 24180 $ & $ 788.4 \pm 9.4 $ & $ 19920 $ & $ 122.88 $ & $ 1.22 \pm 0.12 $ \\
        $ 10.101 $ & $ 16980 $ & $ 544.2 \pm 6.5 $ & $ 148860 $ & $ 117.45 $ &$ 4.35 \pm 0.41 $ \\
        $ 10.634  $ & $ 11220 $ & $ 817.4 \pm 9.8 $ & $7800  $& $ 128.41 $& $ 15.8 \pm 1.5 $ \\
        $ 12.228 $ & $ 63380 $ & $ 639.7 \pm 7.7$ & $ 9120 $ & $ 123.09 $ & $ 48.1 \pm 4.4$ \\
        $ 13.801 $ & $14700 $ & $ 522.9 \pm 6.2 $& $7860 $ & $ 119.59 $ & $ 125.0 \pm 11.0$ \\
        $ 15.383 $ & $ 3600 $ & $308.8 \pm 3.7$ & $15180  $& $ 118.07 $& $ 163.0 \pm 15.0$ \\
        $ 16.977 $ & $ 8760$ & $ 193.9 \pm 2.3 $ & $ 336120 $ & $ 116.81 $ & $ 112.0 \pm 10.0$ \\
        $ 18.562 $ & $ 3900$ & $ 96.0 \pm 1.2 $ & $ 24300 $ &$ 127.45 $&$128.0 \pm 12.0 $\\
        $ 20.678 $ & $1860 $ & $ 39.9 \pm 0.5 $ & $26940 $ & $ 124.49 $ & $ 72.5 \pm 6.6 $ \\
        \bottomrule 
    \end{tabular}
    \caption{Measurement data for the available beam energies at METAS. $E_{beam}$ is the electron beam energy, $t_{irr}$ the duration of the irradiation, $Q$ is the total electron charge, $t_{dec}$ is the time between the end of irradiation and the HPGe measurement, $m_{Au}$ is the target foil's weight and $A$ is the measured \isotope[196]{Au} activity. The uncertainties of the beam energy and the mass are \SI{25}{\kilo\electronvolt} and \SI{0.01}{\milli\gram} respectively, whereas the uncertainty of the irradiation and decay times are negligible.}  \label{t:measurements}
\end{table}

\subsection{Monte Carlo simulations}

\begin{figure}
  \centering 
  \includegraphics[trim=6cm 3cm 13cm 0cm, clip, width=0.4\textwidth]{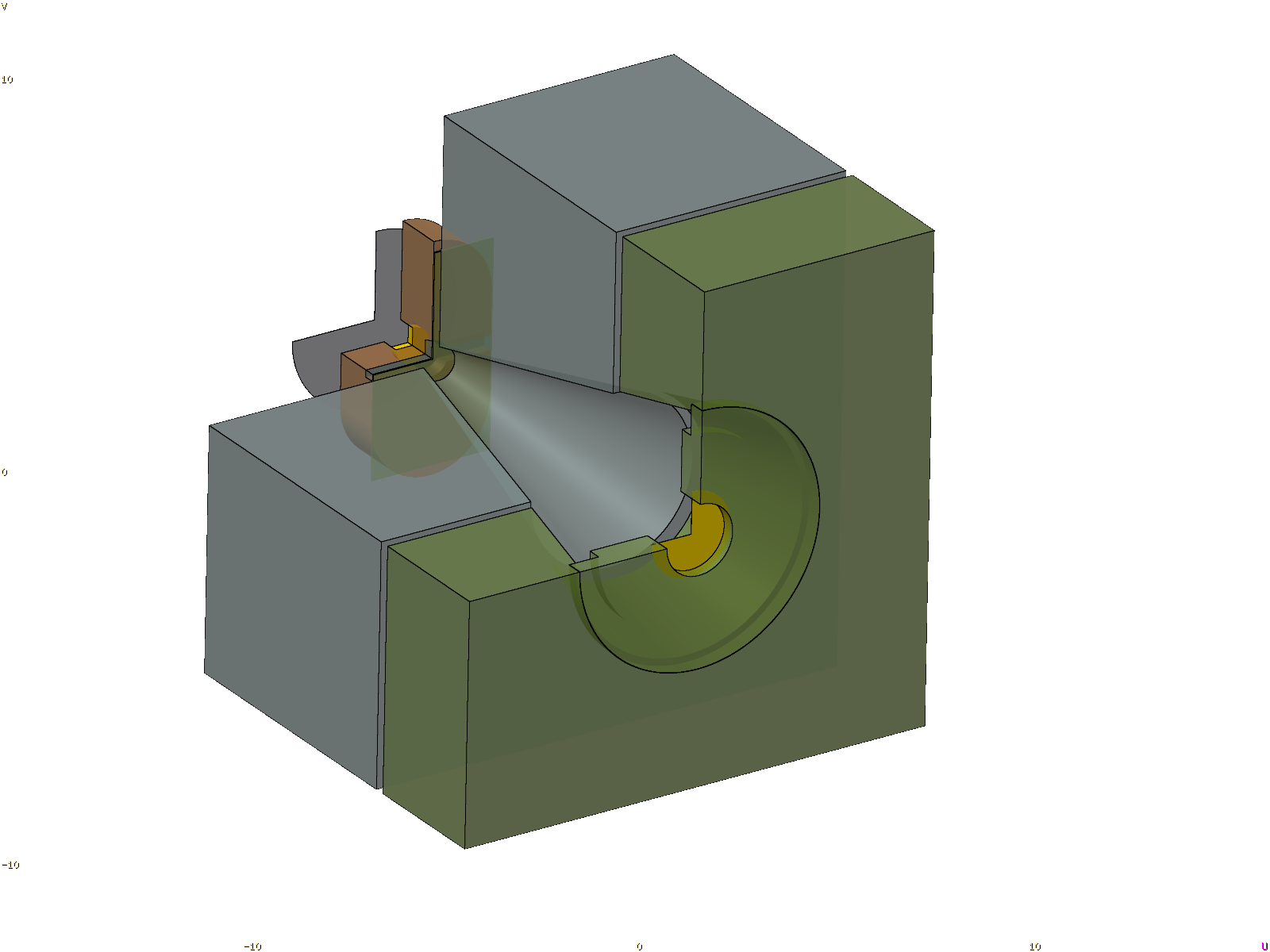}
	\hfill 
  \includegraphics[width=0.52\textwidth]{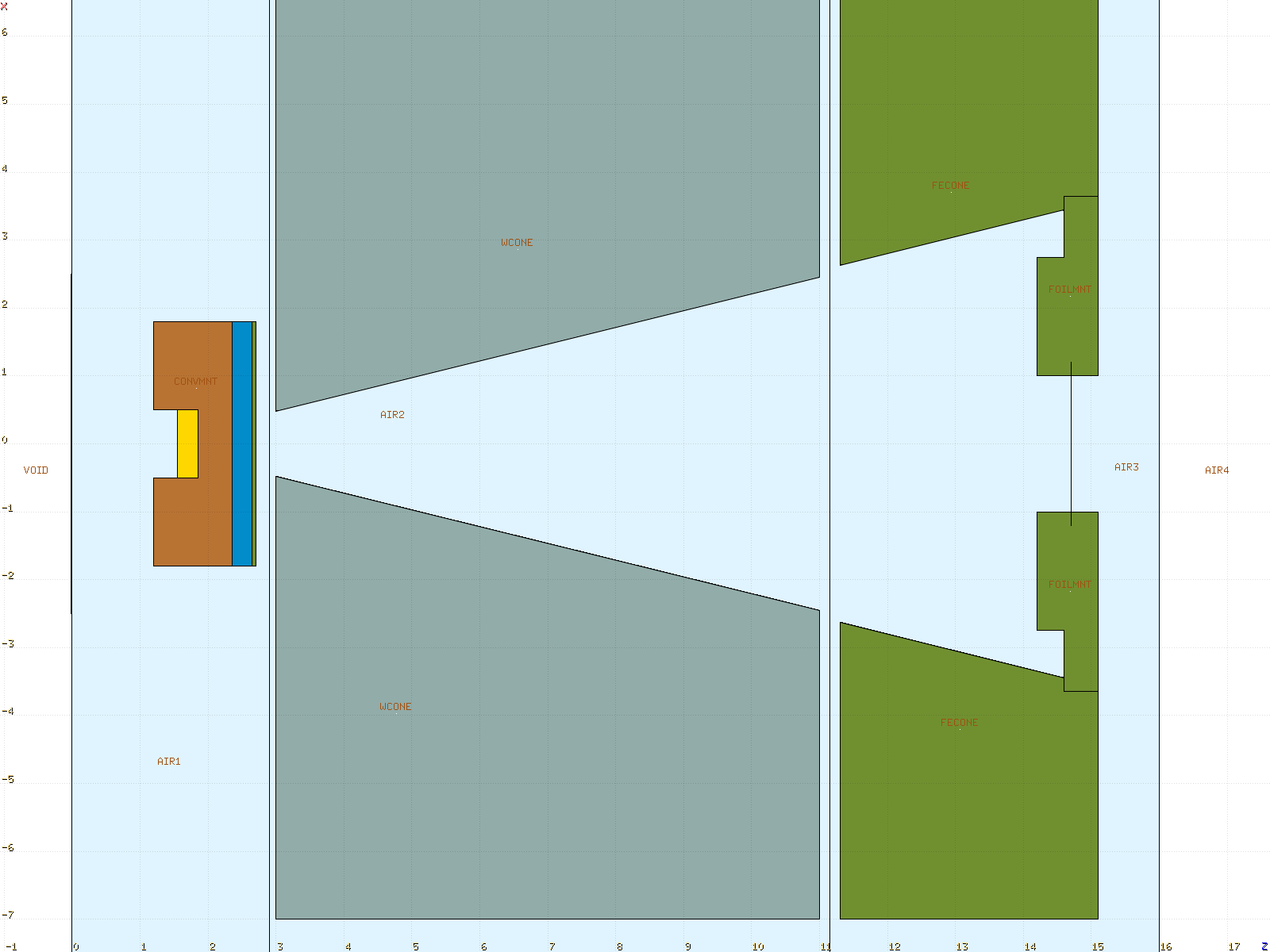}
  \caption{FLUKA implementation of the accelerator head at METAS. } \label{f:accelerator_fluka}
\end{figure}

A key ingredient for measuring photonuclear cross sections is the precise characterization of the photon beam. Nowadays, the gold standard for assessing particle fluences in accelerator environments are Monte Carlo (MC) particle transport simulations of the relevant accelerator elements. In our case, this means simulating the accelerator head with the converter, collimators and target assembly. We implemented the accelerator head's geometry in FLUKA version 4.0 and Flair 3.1 \cite{fluka1, fluka2, fluka3, flair} and independently in Geant4 \cite{geant1,geant2,geant3} based on the technical drawings of the treatment head. Fig.~\ref{f:accelerator_fluka} shows Flair's rendering of the accelerator head with the vacuum window, converter with mount and cooling channels, collimators and target. The full beam line is not part of the simulation as it is not relevant for our purposes. The initial electron beam of the simulation starts in the vacuum pipe, just before going through the vacuum window. The electron beam shape is Gaussian in the two directions perpendicular to the beam axis with a FWHM of $ 3 \, \mathrm{mm}$. Also the beam energy profile is implemented as Gaussian with a FWHM of $25 \, \mathrm{keV}$. 

\begin{figure}
  \centering 
  \includegraphics[width=0.7\textwidth]{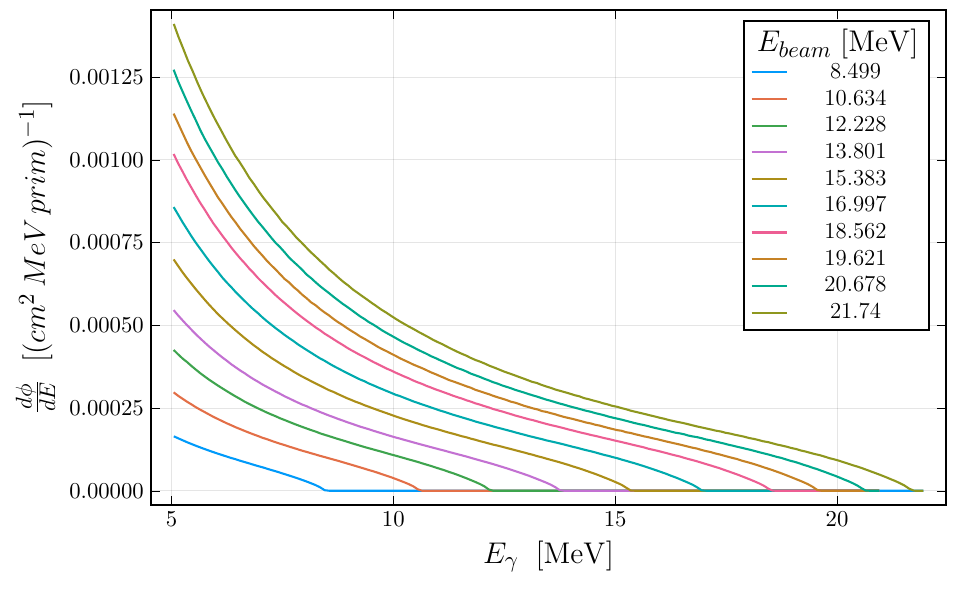}
  \caption{Photon fluence per primary at the irradiation point from FLUKA for various beam energies.}\label{f:fluka_fluence}
\end{figure}

Fig.~\ref{f:fluka_fluence} shows the differential photon fluence for different electron beam energies. Due to the accelerator setup, a standard Bremsstrahlungs fluence spectrum is expected. The simulations run for a sufficient number of primaries in order to keep the statistical error on the differential photon fluence well below $1 \%$ even for photon energies close to $E_{beam}$. 

\begin{figure}
  \centering 
  \includegraphics[width=0.47\textwidth]{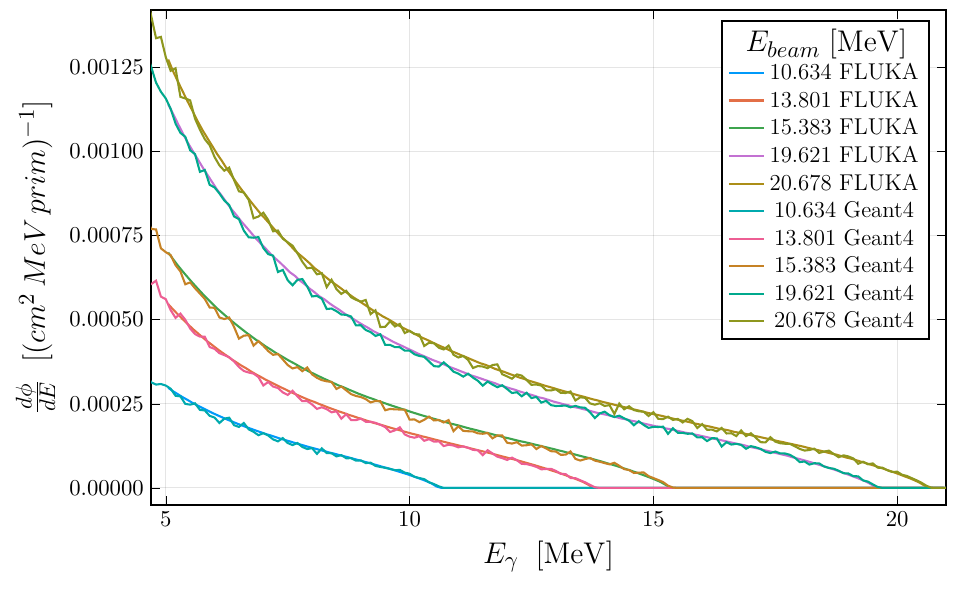}
  \hfill
  \includegraphics[width=0.47\textwidth]{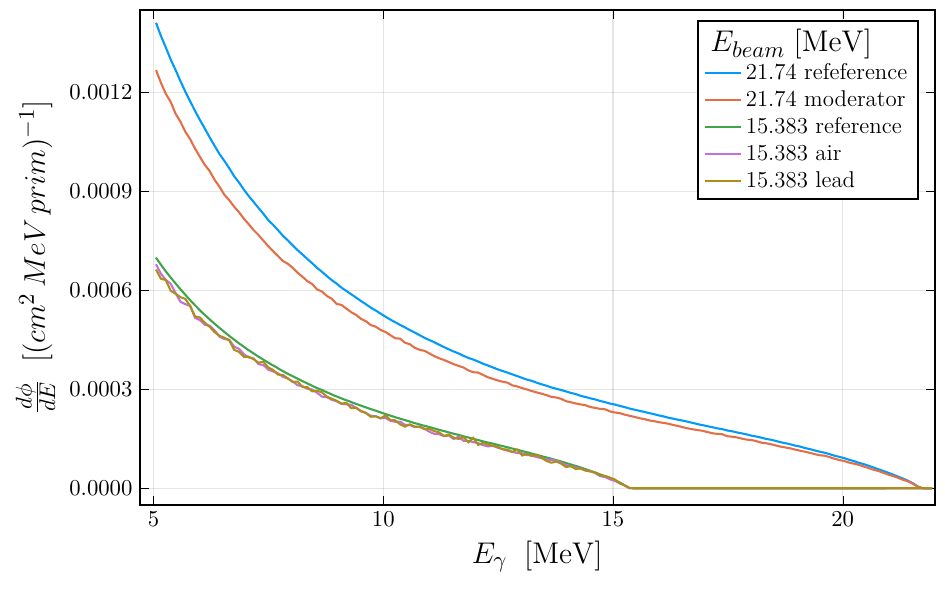}
  \caption{Comparison between the photon fluence from Geant4 and FLUKA (left) and impact of geometry alterations in the FLUKA simulations (right).  }\label{f:geant4_fluka_fluence}
\end{figure}

A thorough assessment of the photon fluence's uncertainty is necessary. First, the left plot in Fig.~\ref{f:geant4_fluka_fluence} shows that the agreement between the Geant4 and FLUKA simulation is very good within the statistical uncertainty (the Geant4 has a slightly higher statistical noise). This means that differences in the material definitions and in the physics implementation of electromagnetic interactions in the two codes have a negligible impact on the differential photon fluence at the target location. Second, we checked that in order to alter significantly the photon fluence above $5 \, \mathrm{MeV}$, rather large changes in the geometry of the accelerator head would be necessary. Adding a $3\, \mathrm{cm}$ thick polyethylene neutron moderator in front of the target merely leads to an overall reduction of $\approx 6\% $ for $E_{beam} = 21.74 \, \mathrm{MeV}$ in the photon fluence at the target, as can be seen in the right plot of Fig.~\ref{f:geant4_fluka_fluence}. Also replacing the water in the cooling cavity of the converter with air or placing a $1 \, \mathrm{mm}$ thick lead foil directly in the photon beam between the converter and the target does not affect significantly the photon fluence (see right plot of Fig.~\ref{f:geant4_fluka_fluence}). 

We also verified the simulation results experimentally with dosimetric measurements. Since the accelerator at METAS is mainly used for metrology, we performed measurements with two calibrated PTW 31014 ionization chambers inside a water phantom. The reference chamber was placed at $100 \, \mathrm{cm}$ from the converter directly on the central photon beam axis and the second chamber was located $ 7.6 \, \mathrm{cm}$ behind. This allowed us to verify the simulation results with two quantities: the absolute dose in the reference chamber, which provides a benchmark for the normalization of the simulation, and the ratio of the doses deposited in the two chambers, which is independent of the normalization or charge measurement and contains information about the lower energy part of the photon spectrum. 

\begin{table}
  \centering
  \begin{tabular}[]{lcc}
    \toprule 
     & $ D_{ref} \, [\mathrm{Gy}/\mathrm{prim}]$ & $ D_{ref}/D_{back} $ \\
    \midrule 
    FLUKA & $ (8.02 \pm 0.29) \cdot 10^{-16}$ &  $1.45 \pm 0.08$  \\
    PTW chamber & $ (8.68 \pm 0.44) \cdot 10^{-16}$ & $ 1.43 \pm 0.10 $ \\
    \bottomrule
  \end{tabular}
  \caption{Comparison of the absolute dose and the ratio of dose to water in two locations between FLUKA simulations and the measured doses in PTW ionzation chambers. }\label{t:dose_measurements}
\end{table}

For an electron beam energy of $15.383 \, \mathrm{MeV}$, Tab.~\ref{t:dose_measurements} reports the dose in the reference chamber and the ratio of the doses from the two chambers. The measurements were repeated five times, which yielded a statistical error of about $3\%$. Furthermore, we assumed an experimental uncertainty on the beam charge measurement of \SI{1.2}{\percent}.  

The FLUKA simulation of this dosimetric measurement involved a significant extension of the accelerator geometry. A flattening filter, additional collimators and the water tank with the ionization chambers, all of which are located after the target for the cross section measurements, had to be implemented. The doses are scored in regions of the size of the chambers according to the PTW specifications (using regional USRBIN scoring). We show the results with the statistical uncertainty of the simulation in Tab.~\ref{t:dose_measurements}. The simulation results agree very well with the measurements. Despite the fact that with this experimental setup we probe primarily the lower end of the photon energy spectrum, this test measurement gives us confidence that our characterization of the photon beam with FLUKA is accurate. 

The MC simulations do not only provide the differential photon fluence as the input for the cross section measurement, they also provide the yield of the activation products in the target material. This is a valuable benchmark and comparison for the experimental determination of the yield. The activation products are scored with RESNUCLE card with radioactive decay set in semi-analogue mode. Of course, the photonuclear interactions need to be turned on in FLUKA with the PHOTONUC card set to ELECTNUC (we do not need any photonuclear interactions with muons) and with the COALESCE and EVAPORAT settings of the PHYSICS card. For the target material, the same biasing factor applies as in the case of the aforementioned converter biasing. 

Note that FLUKA has its own implementation of photonuclear cross sections. As described in Refs.~\cite{Fasso1997,Fasso2005} FLUKA has its own cross section library for photonuclear interactions for about 190 stable nuclides. In the energy range around the giant dipole resonance (GDR), which is relevant for our study, the photonuclear cross sections are based on an evaluated parametrization done by the FLUKA developers based on available experimental data and theoretical considerations \cite{Fasso2005}. An assessment of the systematic uncertainty on the yield of activation products in FLUKA, in particular \isotope[196]{Au}, would go beyond the scope of this work since it would mean to evaluate the accuracy of the implemented cross section. We therefore only report the statistical uncertainty on the yield from FLUKA.

\subsection{Data analysis}

From the target irradiations at METAS, we obtained the yield of \isotope[196]{Au} in the target foil, the foil's weight and the time integrated electron beam charge. In order to keep all measured information separate from the modelled and/or simulation data, we normalized the measured yield of \isotope[196]{Au} to the number of primary particles and unit volume and denote it as $y_{Au} (E_{beam})$. This involves the decay correction for the time between the irradiation of the target and its HPGe measurement. We assume a decay constant for \isotope[196]{Au} of $\lambda = (1.3009 \pm 1.27 \cdot 10^{-4})\cdot 10^{-6} \, \mathrm{s^{-1}}$ according to Ref.~\cite{xiaolong2007}.

The relation of $y_{Au} (E_{beam})$ to the production cross section and photon fluence $\phi$ is 

\begin{equation}\label{eq:yield}
  y_{Au} (E_{beam}) \eq \rho_{Au} \, \int_{E_{th}}^{E_{beam}} dE' \, \frac{d\phi (E_{beam},E')}{dE'} \cdot \sigma(E') \;,
\end{equation}
where $E_{th}$ is the threshold energy, $\rho_{Au}$ is the number density of target nuclei, and $\sigma$ is the photonuclear cross section for $\isotope[197]{Au}(\gamma, n)\isotope[196]{Au}$. With the differential photon fluence determined through the FLUKA simulation, it is possible to extract the cross section $\sigma$ from the measured yields $y_{Au} (E_{beam})$. 

The limited number of available electron beam energies implies several restrictions. On the one hand, we had to restrict the shape of $\sigma$ to a truncated Breit-Wigner function

\begin{equation}\label{eq:breitwigner}
  \sigma(E) \eq  \frac{n \cdot k \cdot \Theta(E - E_{th})}{(E^2 - m^2)^2 + m^2 \, \Gamma^2} \;, \qquad \mbox{with} \qquad k \eq \frac{2\sqrt{2}}{\pi} \frac{ \, m \cdot \Gamma \, \sqrt{m^2 (m^2 + \Gamma^2)}}{ \sqrt{m^2 + \sqrt{m^2 (m^2 + \Gamma^2)}}} \;. 
\end{equation}

The threshold energy $E_{th}$ for the $\isotope[197]{Au}(\gamma, n)\isotope[196]{Au}$ reaction is fixed to $8.070 \pm 0.003 \, \mathrm{MeV}$ according to Ref.~\cite{Audi2003}. $n$ is a normalization constant which is fitted to the measured data together with the mass $m$ and width $\Gamma$ of the GDR. 

At energies around the GDR, this is an appropriate model for the cross section in the case of gold. Note that for non-spherical nuclei the cross section might be a combination of two Breit-Wigner functions and also the $\sigma \propto \sqrt{E-E_{th}}$ behavior in the threshold region is not implemented in Eq.~\eqref{eq:breitwigner}. Given the limited number of data points, fitting a more complex parametrization of $\sigma$ is bound to fail. 

We performed a Bayesian fit of the parameters $n$, $m$, and $\Gamma$ using the Turing.jl package \cite{julia,turing}. We believe that the integral in Eq.~\eqref{eq:yield} and the limited number of data points make a Bayesian approach more appropriate. A Gaussian likelihood and the following conservative priors were used for the fitting procedure

\begin{equation}
  \begin{split}
    n & \; \sim \; \mathcal{N}(10^{-24} \, \mathrm{cm^2}, 10^{-24} \, \mathrm{cm^2})  \;, \\
    m & \; \sim \; \mathcal{N}(14.0 \, \mathrm{MeV}, 3.0 \, \mathrm{MeV}) \;,  \\
    \Gamma & \; \sim \; \mathcal{N}(2.0 \, \mathrm{MeV}, 1.0 \, \mathrm{MeV})   \;.
  \end{split}
\end{equation}

In addition, the statistical noise has a normally distributed prior. All of the priors' normal distribution were truncated at $0$. Since the calculation of the posterior distribution requires a large amount of evaluations, the integral in Eq.~\eqref{eq:yield} is performed using the trapezoidal rule. Given the small bin size of the photon fluence of $ 0.1 \, \mathrm{MeV}$ and the smoothness of the integrand function, we assume that the uncertainty on the numerical evaluation of the integral is marginal. 

In our analysis we did not assume any uncertainty on the differential photon fluence from the FLUKA simulations. With the checks described in the previous section, the shape and the normalization of $d\phi / dE$ are well under control. Furthermore, any normalization uncertainty would simply propagate into the parameter $n$ when sampling the posterior distribution. Also, changes in the shape of the photon fluence spectrum, e.g.\ introducing a bin-wise error, is hardly noticeable due to the integration in Eq.~\eqref{eq:yield}.  

From Eq.~\eqref{eq:yield} and the shape of the Breit-Wigner function it is clear that the fit results are mostly dependent on the data points with $E_{beam}$ below the peak of the GDR (see also Ref.~\cite{Nair2008}), i.e.\ in our case $E_{GDR} = 13.7 \, \mathrm{MeV} $ \cite{xiaolong2007}. It was therefore our primary goal to get as much data points in this energy range in order to improve the fit as much as possible.

\section{Results}\label{s:results}

In Tab.~\ref{t:measurements} we present the measured \isotope[196]{Au} yields together with the irradiation data. The data was taken with different target foils and therefore the target's mass was added. Due to the long irradiation time and long half-life of \isotope[196]{Au}, it is safe to assume a negligible uncertainty on $t_{irr}$ and $t_{dec}$. The uncertainty on the target mass measurement is also negligible. The electron beam energy is restricted to a narrow energy range due to the stability criterion of the accelerator design. Additionally, beam energies for individual orbits were determined during commissioning of the accelerator using a magnetic spectrometer (see Fig.~\ref{f:scheme_Microtron}). Here, the energy spread of the electron beam was also measured to be \SI{25}{\kilo\electronvolt}. A statistical uncertainty on the number of counts measured with the HPGe was considered for the \isotope[196]{Au} activity(Tab.~\ref{t:measurements}). 

\begin{figure}[ht]
    \centering
    \includegraphics[width=0.8\textwidth]{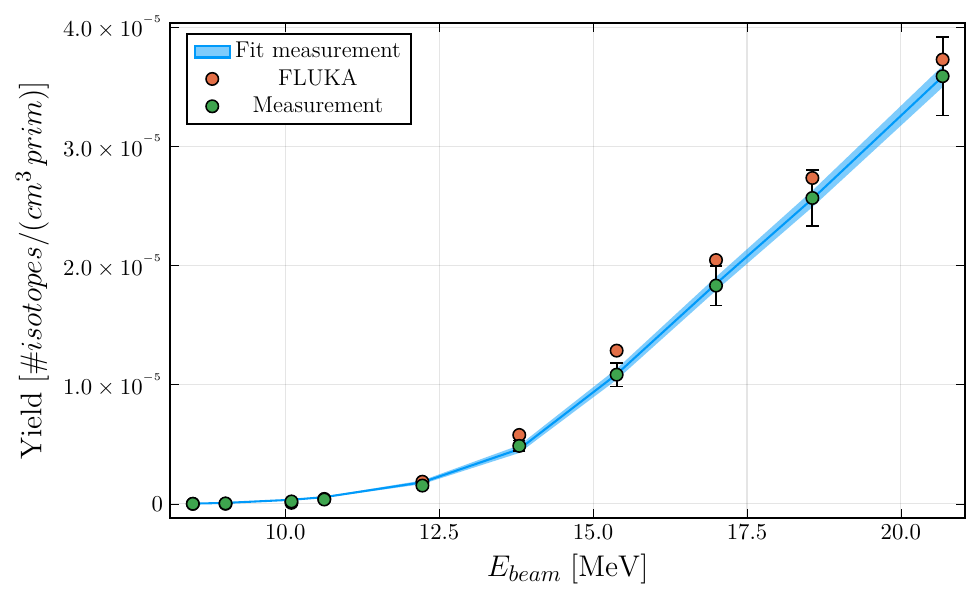}
    \caption{Comparison of the \isotope[196]{Au} yield per primary beam particle, i.e.\ electrons, between FLUKA and the measured data. The blue band shows the yield predicted by the measurement's cross section fit.}\label{f:yield}
\end{figure}

Fig.~\ref{f:yield} shows the decay-corrected and normalized yield from the measurements in comparison with the simulated yield. The uncertainty on the decay constant of \isotope[196]{Au}, retrieved from Ref.~\cite{nudat3}, is negligible. Only at lower energies, close to the $\isotope[197]{Au}(\gamma, n)\isotope[196]{Au}$ reaction threshold, the measured and the simulated yields do not agree well. On the one hand, the measurements in this regime are plagued by long irradiation times and low activities. On the other hand, FLUKA has its own evaluated cross section library and the implementation of the $\isotope[197]{Au}(\gamma, n)\isotope[196]{Au}$ reaction threshold is not disclosed. Note that standard evaluated cross section libraries like TENDL \cite{TENDL2019} and IAEA \cite{kawano2020iaea} differ at threshold energies. 

The error bars on the FLUKA yield in Fig.~\ref{f:yield} are hardly visible. The statistical errors are $7.8\, \%$ and $3.8 \, \%$ for the two lowest beam energies and well below $1\, \% $ for the higher beam energies. The measured yield's error is given only by the uncertainty of the HPGe measurement and is around $10\, \%$ (see also the Tab.~\ref{t:measurements}). The fit prediction from the measurement data is also shown in Fig.~\ref{f:yield}.  

Tab.~\ref{t:fit_results} shows the fit results for the measured and simulated yield, respectively. Clearly, the uncertainties on the parameters of the Breit-Wigner function are relatively small. 

\begin{table}
  \centering
  \begin{tabular}[]{lcccc}
    \toprule 
     & $ n \, [10^{-24} \, \mathrm{cm}^2]$ & $ m \, [\mathrm{MeV}] $ & $ \Gamma \, [\mathrm{MeV}] $ & $ \varepsilon \, [\#iso/(cm^3 prim)] $ \\
    \midrule 
    FLUKA & $2.43 \pm 0.04 $ & $13.5 \pm 0.1 $ & $2.12 \pm 0.20$ & $ (1.70 \pm 0.60) \cdot 10^{-7}$ \\
    Measurement & $2.58 \pm 0.05$ & $14.2 \pm 0.1 $ & $2.76 \pm 0.27$ & $(2.28 \pm 0.81) \cdot 10^{-7}$ \\
    \bottomrule
  \end{tabular}
  \caption{Results from fitting the Breit-Wigner function of Eq.~\eqref{eq:breitwigner} to either the FLUKA or measured \isotope[196]{Au} yield. The parameter's pdf is shown in Fig.~\ref{f:prior_posterior_pdf}. }\label{t:fit_results}
\end{table}

\begin{figure}
    \centering
    \includegraphics[width=0.8\textwidth]{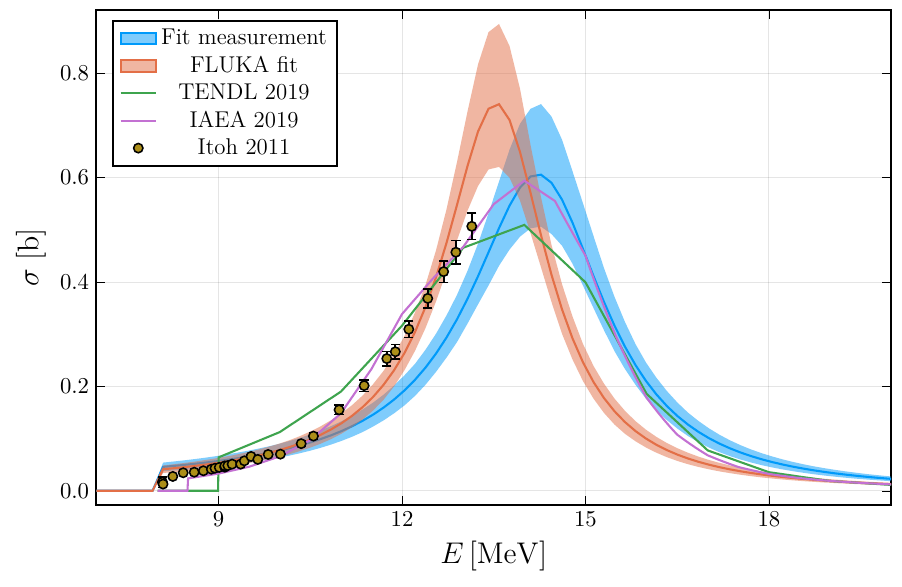}
    \caption{$\isotope[197]{Au}(\gamma, n)\isotope[196]{Au}$ cross section from the measured data fit (see Tab.~\ref{t:fit_results}) in comparison with the cross sections TENDL 2019\cite{TENDL2019}, IAEA 2019\cite{kawano2020iaea} and Ref.~\cite{itoh2011}.}
    \label{f:cross_section}
\end{figure}

Finally, Fig.~\ref{f:cross_section} shows the resulting $\isotope[197]{Au}(\gamma, n)\isotope[196]{Au}$ cross section from the measured yield. The peak of the cross section from the measurement is slightly shifted towards higher energies in comparison with the fitted cross section from FLUKA (see also parameter $m$ in Tab.~\ref{t:fit_results}). However, the evaluated cross sections from Refs.~\cite{TENDL2019,kawano2020iaea} have the peak well within the error band of the measurement fit.  

\section{Discussion}

\begin{figure}
    \centering
    \includegraphics[width=\textwidth]{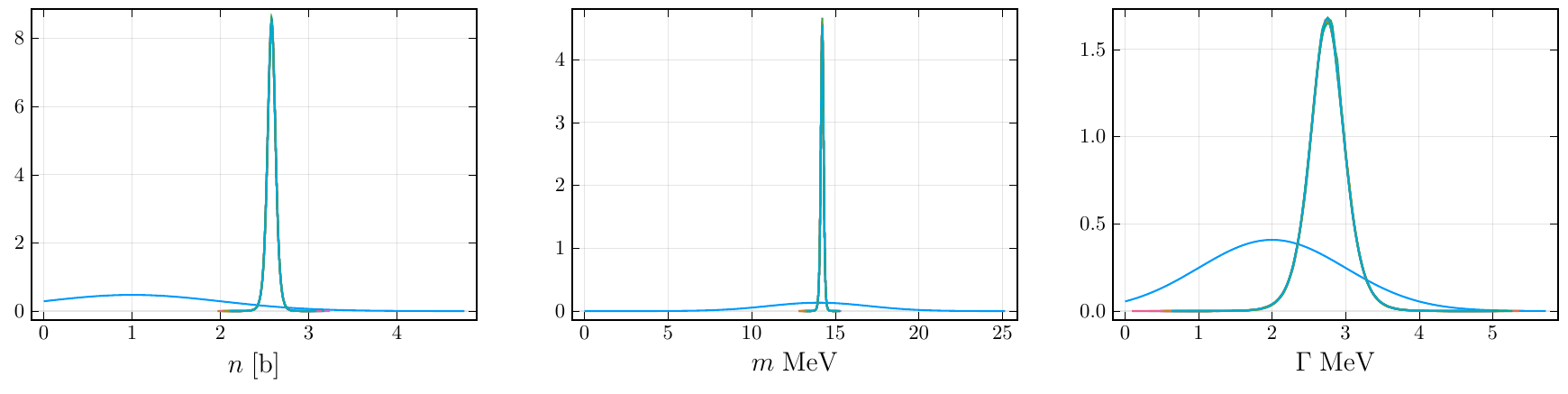}
    \caption{Comparison of the prior (blue) and posterior (green) pdf for the three fitting parameters $n$, $m$ and $\Gamma$. The measurement data constrains the parameter rather strongly.  }
    \label{f:prior_posterior_pdf}
\end{figure}

The results from fitting a Breit-Wigner curve to the measured \isotope[196]{Au} yield in Tab.~\ref{t:fit_results} show that the methodology discussed in Sec.~\ref{s:methods} allows to determine photonuclear cross sections with the electron accelerator at METAS. For the reference process $\isotope[197]{Au}(\gamma, n)\isotope[196]{Au}$, the uncertainties on the predicted cross section in Fig.~\ref{f:cross_section} are well within the range of the experimental and evaluated cross sections from the literature. Even with the restricted number of data points, i.e.~available beam energies $E_{beam}$ in Tab.~\ref{t:measurements}, the fitting parameters are strongly constraint. Fig.~\ref{f:prior_posterior_pdf} demonstrates how the fit strongly reduced the width of the posterior pdf compared to the priors. Therefore there is a significant gain in information over our prior knowledge.

Unsurprisingly, the fit is mostly sensitive to the yield values around the resonance peak. In this region the Breit-Wigner curve has the strongest gradient and the photon fluence is still large. Therefore, slight changes in the upper integration limit in Eq.~\eqref{eq:yield} have a larger impact on the yield. 

Given the multiple orders of magnitude of the measured and simulated yields, it might be tempting to perform the fit in log space. The $\log (y_{Au})$ has a high gradient at low energies and flattens towards $\sim 20 \, \mathrm{MeV} $. Fitting the log of the yield therefore gives a higher weight to the lower energy data points. The measured points in the threshold region are, however, affected by low count statistics and therefore may be affected by systematic uncertainties. Furthermore the convergence of the fit worsens in log space and the relative errors on $n$, $m$ and $\Gamma$ increase compared to the results in Tab.~\ref{t:fit_results}. 

Comparing the fits to the measured and simulated yields, it is clear that there is an underestimation of the yield starting around $11 \, \mathrm{MeV} $. This is the reason why the FLUKA data lead to a higher peak of the Breit-Wigner curve (see Fig.~\ref{f:cross_section}). Interestingly, these lower values of the yield do not drive $m$ to higher energies. This is due to the fact that close to the threshold energy, the simulated yield is lower compared to the measured one.   

\begin{figure}
    \centering
    \includegraphics[width=0.95\textwidth]{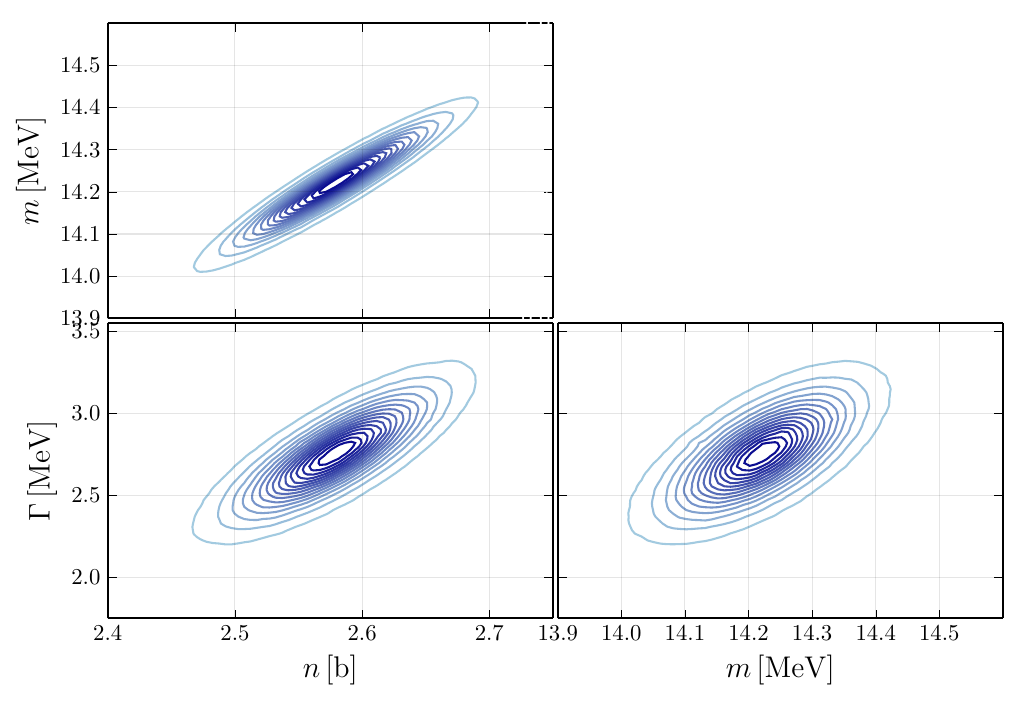}
    \caption{Pair plot of the three parameters for the fit to the measured yield showing a strong correlation of the posterior distributions. }
    \label{f:pair_plot}
\end{figure}

Fig.~\ref{f:pair_plot} presents the pair plot of the fitting parameters for the case of the measured data. Clearly, the parameters $n$, $m$ and $\Gamma$ are strongly correlated among each other. This stems from the integration in Eq.~\ref{eq:yield} since it averages out the Breit-Wigner curve. Increasing the normalization $n$ will drive $m$ towards higher energies in order to decrease the overlap between the peak of the Breit-Wigner curve and the high fluence region. The same reasoning works for the other two correlations, i.e.\ higher values of $m$ require a wider cross section that is able to still catch higher fluence contributions at lower energies and increasing $n$ makes the Breit-Wigner function flatter. 

 The number of beam energies, of course, limits the goodness of the fit. As a cross check, we simulated more beam energies and performed a fit of the resulting yield. This means we effectively reverse engineered the cross section implemented in FLUKA. With a total of $18$ beam energies, of which $12$ lied below $14 \, \mathrm{MeV}$, the parameters are constrained much stronger. Even for $\Gamma$ the relative error drops below $1 \, \%$. This shows that despite the averaging of the integral in Eq.~\eqref{eq:yield} the method can improve with more data points. The degeneracies of the parameters from Fig.~\ref{f:pair_plot} remain. 

Despite the good results for the $\isotope[197]{Au}(\gamma, n)\isotope[196]{Au}$ cross section, our method and experimental setup face some challenges. On one side there is a model dependence in the sense that we need to rely on the assumption of a Breit-Wigner shape of the cross section. This assumption should be questioned in particular when measuring cross sections with non-spherical target nuclei that require more complex fitting functions. Increasing the number of fitting parameters, such as e.g.\ if the sum of two Breit-Wigner functions, would certainly require additional data points in order to keep the uncertainties at an acceptable level. On the other side, our method requires input from MC simulations which could be viewed as a limitation or, at least, as a source of systematic uncertainties. A good characterization of the experimental setup and verification of the MC simulations (see also Sec.~\ref{s:methods}) is there fore key for a successful determination of photonuclear reaction cross sections. 

\section{Conclusions and outlook}

In this study we showed that it is possible to measure the photonuclear reaction cross section for the reaction $\isotope[197]{Au}(\gamma, n)\isotope[196]{Au}$ using the Microtron at METAS. The \isotope[196]{Au} yield in thin gold foils is determined by irradiating thin gold foils with photons and measuring the induced activity with a HPGe spectrometer. Assuming that the cross section follows a Breit-Wigner curve and with the modelling of the photon fluence using MC simulations, we were able to reproduce the reference values for the $\isotope[197]{Au}(\gamma,n)\isotope[196]{Au}$ cross section through a Bayesian fitting procedure. Even with a limited number of beam energies, the Bayesian fitting procedure yields low uncertainties on the parameters of the Breit-Wigner shape of the cross section. Our results crucially rely on an accurate characterization of the photon fluence spectrum as well as on the precise determination of the induced activity and the electron beam current. The method presented in this study can be translated easily to other photonuclear reactions, for which the cross sections are hardly known. Depending on the process under investigation, more fitting parameters will be required (sum of two Breit-Wigner functions or multi-isotopic target materials). Therefore, we envision that more beam energies would be required for more complex fits to the data. 
The isotope \isotope[226]{Ra} is an intriguing candidate. The cross-section for this reaction would be interesting, considering large-scale production of \isotope[225]{Ac} for targeted alpha therapy using the photonuclear route.
With the presented methodology we laid the groundwork to accurately measure photonuclear cross sections with a relatively simple setup. The method can be easily applied at other facilities which might have access to higher beam energies or higher beam currents. 

In sum, our study not only validates the methodology for measuring the photonuclear reaction cross section using the  $\isotope[197]{Au}(\gamma, n)\isotope[196]{Au}$ reaction, but also opens up new avenues for extending this approach to other isotopes and applications.

\section*{Funding}
This study is supported by Swiss National Science Foundation Sinergia grant PHOtonuclear Reactions (PHOR): breakthrough research in radionuclides for theranostics awarded to A. T\"urler, S.Braccini and C. Kottler; Schweizerischer Nationalfonds zur Förderung der Wissenschaftlichen Forschung (CRSII5\_180352).

\section*{CRediT authorship contribution statement}
\textbf{Saverio Braccini:} Supervision, Project administration, Funding acquisition, Writing - Review \& Editing,
\textbf{Pierluigi~Casolaro:} Conceptualization, Methodology, Software, Validation, Investigation, Writing - Original Draft, Writing - Review \& Editing, Visualization, Formal analysis, Data Curation,
\textbf{Gaia Dellepiane:} Investigation, Writing - Review \& Editing,
\textbf{Christian Kottler:} Supervision, Project administration, Funding acquisition, Writing - Review \& Editing,
\textbf{Matthias L\"uthi:} Conceptualization, Methodology, Software, Validation, Investigation, Writing - Original Draft, Writing - Review \& Editing, Visualization, Formal analysis, Data Curation,
\textbf{Lorenzo Mercolli:} Methodology, Software, Validation, Formal analysis, Investigation, Data Curation, Writing - Original Draft, Writing - Review \& Editing, Visualization,
\textbf{Peter Peier:} Resources, Writing - Review \& Editing, Supervision,
\textbf{Paola Scampoli:} Supervision,
\textbf{Andreas T\"urler:} Writing - Review \& Editing, Project administration, Funding acquisition.

\newpage

\bibliography{mybibfile.bib}

\end{document}